\documentclass[aps,preprint,nofootinbib,groupedaddress,showpacs]{revtex4}

\usepackage{graphics}
\usepackage{graphicx}
\usepackage{amssymb}
\usepackage{amsmath}
\usepackage{amsfonts}

\begin{document}

\title{Stress Tensor of the Hydrogen Molecular Ion}

\author{Kazuhide Ichikawa}
\author{Akitomo Tachibana} 
\email{akitomo@scl.kyoto-u.ac.jp}
\affiliation{Department of Micro Engineering, Kyoto University, Kyoto 606-8501, Japan}

\date{\today}

\begin{abstract}
The electronic stress tensor of the hydrogen molecule ion H$_2^+$ is investigated for the ground state ($\sigma_g$1s) 
 and the first excited state ($\sigma_u^*$1s) 
 using their exact wave functions. A map of its largest eigenvalue and corresponding eigenvector
is shown to be closely related to the nature of chemical bonding. 
For the ground state, we also show the spatial distribution of interaction energy density to describe
in which part of the molecule stabilization and destabilization take place.
\end{abstract}



\maketitle

The stress tensors are used widely for description of internal forces of matter. 
They have been also established in quantum mechanical context \cite{Pauli1980}, although it seems that basic idea dates back to Ref.~\cite{Schrodinger1926}.
Later, applications for quantum systems are investigated in many literatures from various aspects \cite{Epstein1975, Bader1980, Bamzai1981a, Nielsen1983, Nielsen1985, Folland1986a, Folland1986b, Godfrey1988, Filippetti2000, Tachibana2001, Pendas2002, Rogers2002, Tachibana2004, Tachibana2005, Morante2006, Szarek2007, Szarek2008, Tao2008, Szarek2009, Ayers2009}.
For example, Ref.~\cite{Nielsen1983} and followers have focused on stress tensor, which is associated with forces on nucleus.
Our focus in this paper is electronic stress tensors, which measure effects caused by internal forces acting on electrons in
molecules and particularly those between bonded atoms \cite{Tachibana2001,Tachibana2004,Tachibana2005}.
Such electronic stress tensor has been successfully used to define bond orders expressing the strength of chemical bonds \cite{Szarek2007}. 
It should be noted that people sometimes use different definitions for the stress tensor due to their own purpose and formalism.
We use the expression derived by one of the authors in Ref.~\cite{Tachibana2001} which we believe most straightforward and suitable to discuss chemical bonds.

In this paper, we elucidate the role of the electronic stress tensor using hydrogen molecule ion H$_2^+$, the simplest of all molecules, whose exact wave functions are known. 
Although there have been analyses of electronic stress tensor of other molecules using approximate wave functions expanded in the Gaussian functions \cite{Tachibana2001,Tachibana2004,Tachibana2005,Szarek2007,Szarek2008,Szarek2009}, analysis using exact wave functions has not been conducted.
We calculate spatial distributions of stress tensor and other associated quantities from the exact wave functions of H$_2^+$ and examine their detailed features that can be used to illustrate the chemical bond of the molecule. 
We believe that this paper demonstrates usefulness of the electronic stress tensor for an analysis of chemical bonding in the simplest but most exact form. 
\footnote{
Calculations of stress tensor field as regards a H$_2^+$ molecule are found in Refs.~\cite{Bamzai1981b,Godfrey1990} but with different definitions from ours. 
}

We begin by showing our expression of the electronic stress tensor $\tau^{S}_{ij}$ constructed from the H$_2^+$ wave function $\psi$. 
That is,
\begin{eqnarray}
\tau^{S}_{ij} = \frac{\hbar^2}{4m} \left(   \psi^* \partial_i \partial_j \psi - \partial_i \psi^* \partial_j \psi  + c.c. \right), \label{eq:stress}
\end{eqnarray}
where $\{i,j\} = \{1, 2, 3\}$ denote spatial coordinates, $m$ is the electron mass, and $c.c.$ stands for complex conjugate. 
The expression for many-electron systems, that is, for almost every other molecule, is found in  Ref.~\cite{Tachibana2001}.
There, the stress tensor has been derived by field theoretic method which is applicable to many-particle system. Here, for illustrative purpose, we derive it using the Schr\"{o}dinger equation for a single electron in H$_2^+$.

The time-dependent Schr\"{o}dinger equation reads
\begin{eqnarray}
i\hbar \frac{\partial \psi}{\partial t} = -\frac{\hbar^2}{2m} \nabla^2 \psi + V \psi, \label{eq:schrodinger}
\end{eqnarray}
where the potential energy is assumed to be real. It is well-known that the continuity equation $\partial n / \partial t + \boldsymbol{\nabla} \cdot \boldsymbol{j} = 0$ 
holds by defining probability density $n = | \psi |^2$ and probability flux $\boldsymbol{j} = (\hbar/m) {\rm Im} (\psi^* \boldsymbol{\nabla}\psi )$. Now, the equilibrium equation for electronic stress tensor 
is obtained from an equation of motion of $\boldsymbol{j}$. Namely, using Eq.~\eqref{eq:schrodinger}, time derivative of $\boldsymbol{j}$ can be written as 
\begin{eqnarray}
\frac{\partial \boldsymbol{j}}{\partial t} = -\frac{\hbar^2}{4m^2}\left( \boldsymbol{\nabla}\psi \cdot \nabla^2 \psi^*  - \psi^* \boldsymbol{\nabla}\nabla^2\psi + c.c.  \right)  
-\frac{|\psi|^2}{m} \boldsymbol{\nabla} V. \label{eq:djdt} \nonumber
\end{eqnarray}
For the steady state, which is the case of molecular systems we consider, since $\partial \boldsymbol{j} /\partial t = 0$, we obtain the equilibrium equation
\begin{eqnarray}
  \partial_j \tau^{S}_{ij} + F_{L,i} = 0, \label{eq:equilibrium}
\end{eqnarray}
where $F_{L,i} = - |\psi|^2 \partial_i V$ is the Lorentz force generated by nuclei. For later convenience, we here define the first term in the equation as tension $F_{\tau,i} \equiv \partial_j \tau^{S}_{ij}$.
 From this equation, we see that $\tau^{S}_{ij}$ is the stress caused by purely quantum mechanical effect.

\begin{figure}
\begin{center}
 \includegraphics{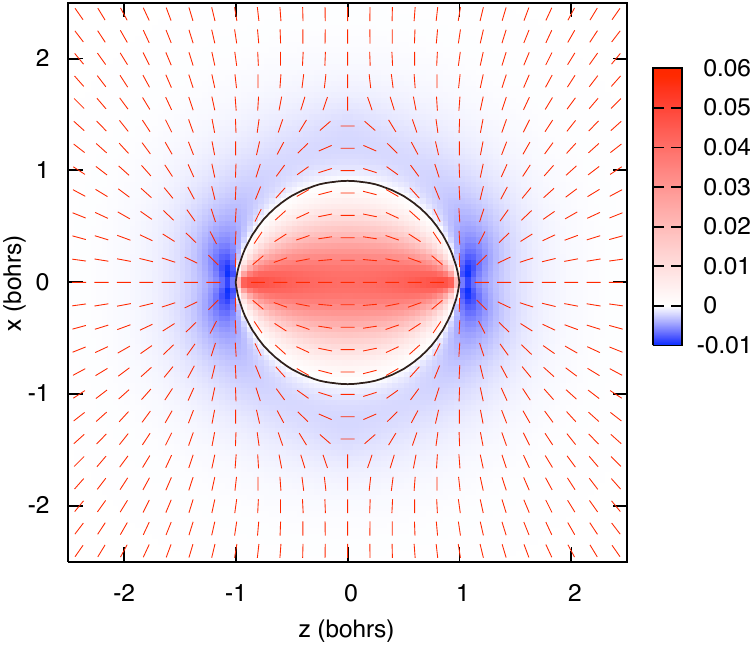}
 \includegraphics{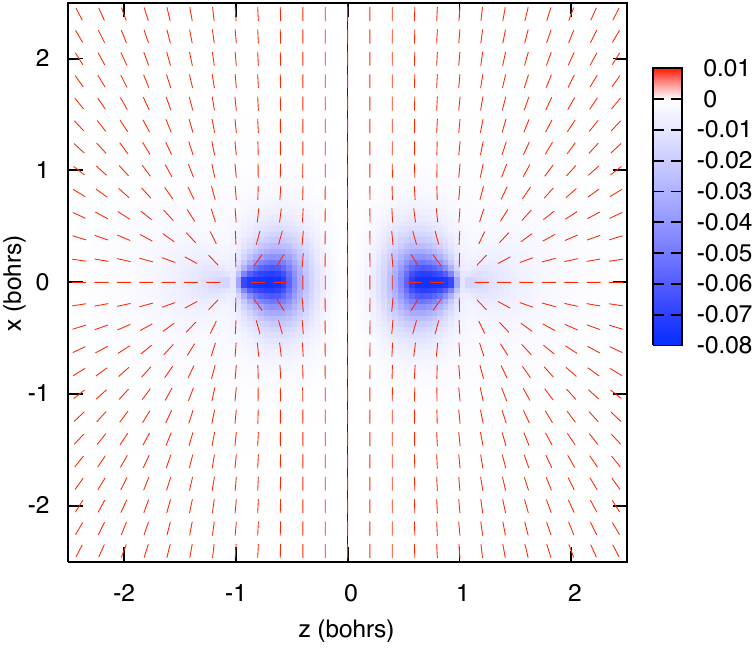}
 \caption{The spatial distributions of the largest eigenvalue and corresponding eigenvectors of stress tensor of H$_2^+$ molecule are plotted in the plane including two H nuclei. They are located at $(z, x) = (-1.0, 0.0)$ and (1.0, 0.0). The left panel is for the ground state ($\sigma_g$1s) and the right panel is for the first excited state ($\sigma_u^*$1s). The black solid line shows the zero surface of the eigenvalue. For $\sigma_g$1s state, it is positive inside the circle and is negative outside. For the $\sigma_u^*$1s state, it is negative on both sides of the surface.}
 \label{fig:eigvec}
\end{center}
\end{figure}

\begin{figure}
\begin{center}
 \includegraphics[width=8cm]{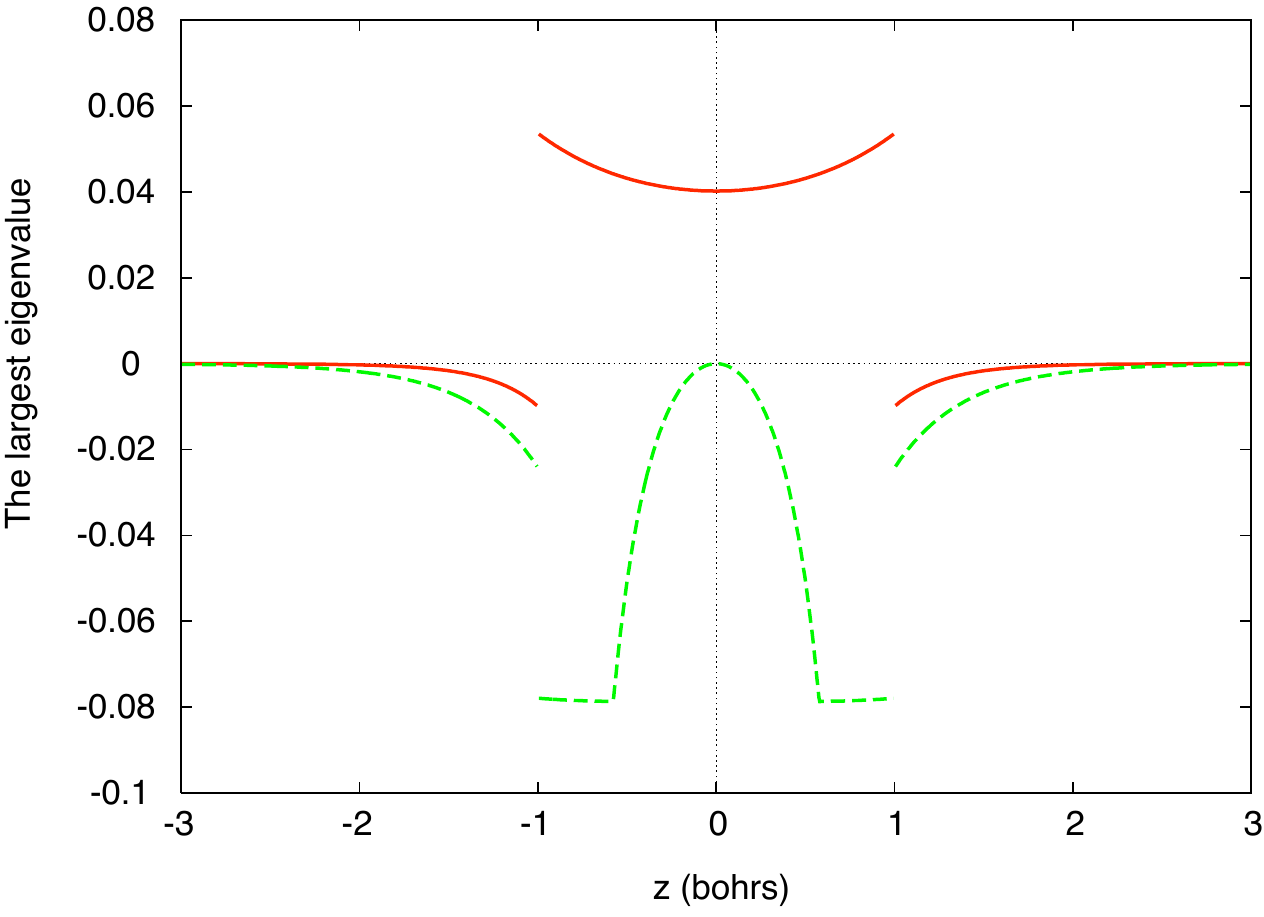}
 \caption{The largest eigenvalue of the stress tensor along the internuclear axis (along the $z$-axis in Fig.~\ref{fig:eigvec}). The red solid line is for the ground state ($\sigma_g$1s) and the green dashed line is for the first excited state ($\sigma_u^*$1s).}
 \label{fig:eig_z}
\end{center}
\end{figure}

We compute the stress tensor defined by Eq.~\eqref{eq:stress} using the exact wave functions of H$_2^+$ \cite{Bates1953}. As is usually done for the stress tensor, we examine its largest eigenvalue and corresponding eigenvector. The sign of the largest eigenvalue tells whether electrons at a certain point in space feel tensile force (positive eigenvalue) or compressive force (negative eigenvalue) and the eigenvector tells direction of the force.

 The case for the ground state ($\sigma_g$1s) is plotted in the plane including two H nuclei as shown in the left panel of Fig.~\ref{fig:eigvec}. The origin of the coordinate is taken to be the midpoint of the two nuclei. The eigenvalue along the internuclear axis is also plotted as red solid line in Fig.~\ref{fig:eig_z}. The internuclear distance is taken to be 2.0 bohrs, which gives the equilibrium distance $R_e$ (to be more precise, $R_e=1.9972$\,bohrs \cite{Schaad1970} but it does not make practical difference in our argument below). The region with positive eigenvalue spreads between the nuclei which corresponds to the formation of a chemical bond. The electronic stress tensor has this useful feature that it can express something is pulled up (that is, tensile force exists) when a covalent bond is formed. In detail, the positive eigenvalue region is bounded by a closed sphere-like surface that touches two H nuclei. Also, in that region, the eigenvectors forms a bundle of flow lines that connects the H nuclei. Such a region, called ``spindle structure" \cite{Tachibana2004}, is clearly visible using the analysis with the exact wave function. 

 The case for the first excited state ($\sigma_u^*$1s) is shown in the right panel of Fig.~\ref{fig:eigvec} on the same plane and with the same internuclear distance. See also the green dashed line in Fig.~\ref{fig:eig_z} for the eigenvalue along the internuclear axis. Since this state does not form a bound state, we do not expect a bond between the H nuclei. Indeed, there is no region with positive eigenvalue between the H nuclei for this case. In addition, eigenvectors are perpendicular to the internuclear axis at the region around the middle of the two H nuclei, showing there is no force connecting them. 

These results, positive eigenvalue for bonding orbital and negative eigenvalue for antibonding orbital, are consistent with Ref.~\cite{Tachibana2004}, where these features are pointed out using H$_2$ and He$_2$, respectively.

\begin{figure}
\begin{center}
 \includegraphics{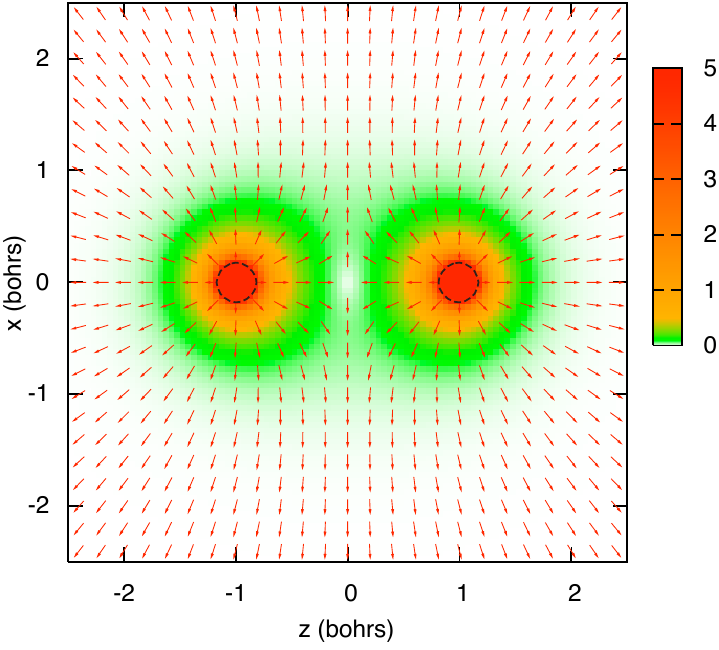}
 \includegraphics{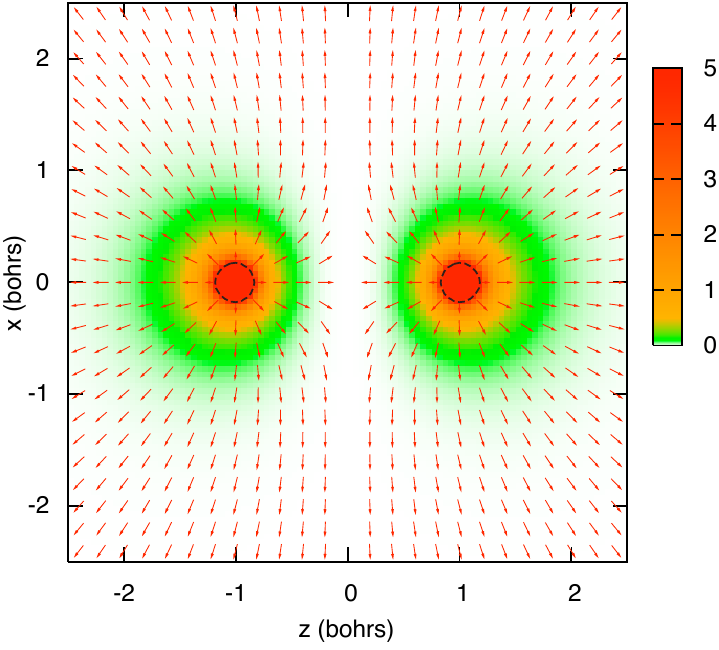}
\end{center}
 \caption{The spatial distributions of the tension $\boldsymbol{F}_\tau$ of H$_2^+$ molecule are plotted similarly to Fig.~\ref{fig:eigvec}. The left panel is for the ground state ($\sigma_g$1s) and the right panel is for the first excited state ($\sigma_u^*$1s). Normalized tension vectors $\boldsymbol{F}_\tau/|\boldsymbol{F}_\tau|$ are shown by arrows and the norm $|\boldsymbol{F}_\tau|$ is depicted by a color map. $|\boldsymbol{F}_\tau|$ diverges at the positions of the nuclei. Contours which correspond to $|\boldsymbol{F}_\tau| = 5$ are drawn by black dashed lines. $|\boldsymbol{F}_\tau|=0$ at (0.0, 0.0) for the $\sigma_g$1s state and on $z=0$ for the $\sigma_u^*$1s state, where arrows are not shown.}
 \label{fig:tension}
\end{figure}

We now examine the tension field $F_{\tau,i} \equiv \partial_j \tau^{S}_{ij}$ of the H$_2^+$ molecule. 
Writing explicitly, 
\begin{eqnarray}
\boldsymbol{F}_\tau = -\frac{\hbar^2}{4m}\left( \boldsymbol{\nabla}\psi \cdot \nabla^2 \psi^*  - \psi^* \boldsymbol{\nabla}\nabla^2\psi + c.c.  \right).  
\end{eqnarray}
Since this force, which cancels the classical Lorentz force at each point in space [Eq.~\eqref{eq:equilibrium}], also expresses purely quantum mechanical effects, it is considered to carry some information of a chemical bond. The spatial distribution of tension vector is plotted for $\sigma_g$1s and $\sigma_u^*$1s states in Fig.~\ref{fig:tension}. On comparing Figs.~\ref{fig:eigvec} and \ref{fig:tension}, we see that in contrast with the case of the eigenvalue and eigenvector of the stress tensor, there is not much difference between the $\sigma_g$1s state and the $\sigma_u^*$1s state for the case of the tension vector. In Fig.~\ref{fig:eigvec}, there is a striking difference between   $\sigma_g$1s and $\sigma_u^*$1s that the former develops a region with positive eigenvalue while the latter exhibit negative region for the most of the space. There is also a marked difference in the pattern of eigenvectors at the internuclear region. In Fig.~\ref{fig:tension}, the two states show a relatively similar pattern of the tension vector field. One difference can be pointed out that there is a point $(x, z)=(0.0, 0.0)$ at which tension vanishes for the $\sigma_g$1s state while the $\sigma_u^*$1s states has a plane $z=0$ on which it vanishes. However, overall feature resembles closely to each other. This indicates that the stress tensor has much more information as regards a chemical bond. 

We note, however, that the tension can provide a handy way to characterize a chemical bond. Actually, it has been used to define an index of bond strength and turned out to be very useful \cite{Szarek2007,Szarek2008,Szarek2009}. In Ref.~\cite{Szarek2007}, it has been proposed to calculate energy density [defined below, Eq.~\eqref{eq:energy}] at the ``Lagrange point," which is defined as the vanishing point of the tension such as we have found at the midpoint of the internuclear axis of the ground state H$_2^+$ molecule in Fig.~\ref{fig:tension}. It should also be noted that the existence of the Lagrange point alone cannot guarantee a chemical bond as is easily seen from the tension of the $\sigma_u^*$1s state of H$_2^+$, which has the Lagrange point (or rather the Lagrange plane $z=0$). Hence, the tension can be a useful tool when combined with analysis of the stress tensor.

\begin{figure}
\begin{center}
 \includegraphics{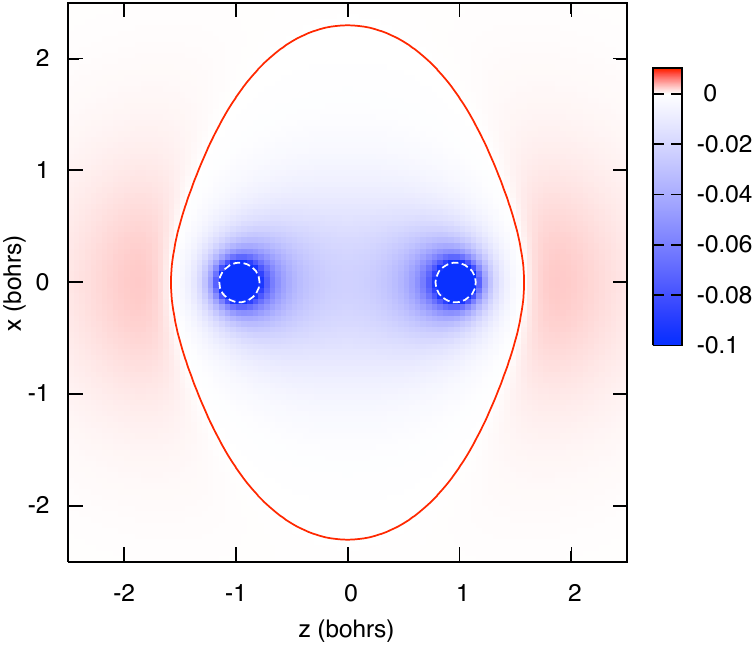}
 \includegraphics{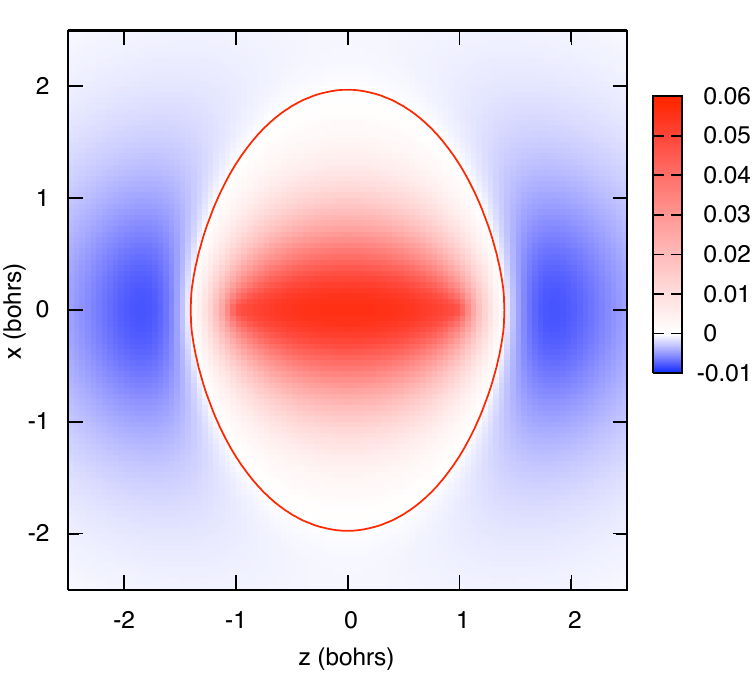}
\end{center}
 \caption{The spatial distributions of the interaction energy density $\Delta \varepsilon_\tau$ (left panel) and the differential electron density $\Delta n$ (right panel) of H$_2^+$ molecule are plotted for the ground state ($\sigma_g$1s). The zero surface is shown by the red solid line for each panel. The interaction energy diverges at the positions of the nuclei and contours that correspond to $\varepsilon_\tau = -0.1$ are shown by white dashed lines in the left panel.}
 \label{fig:int}
\end{figure}

Finally, we describe our definition of energy density, which is derived from the stress tensor, and discuss the interaction energy density of the ground state of the H$_2^+$ molecule. In Ref.~\cite{Tachibana2001}, it has been proposed to define energy density $\varepsilon_\tau$ from the trace of the stress tensor as
\begin{eqnarray}
\varepsilon_\tau &\equiv& \frac{1}{2} \sum_i  \tau^S_{ii} \\
&=& \frac{\hbar^2}{8m} \left(   \psi^* \nabla^2 \psi - \boldsymbol{\nabla} \psi^* \cdot \boldsymbol{\nabla} \psi  + c.c. \right). \label{eq:energy}
\end{eqnarray}
Note that this definition gives correct total energy when integrated over the whole space and the virial theorem is applied. Furthermore, we consider the difference of the energy density between the ground state and the separated atoms. We call this interaction energy density $\Delta \varepsilon_\tau$ \cite{Szarek2007}, and it is a useful quantity to discuss where in a molecule stabilization or destabilization take place by chemical bond formation. 

For the case of H$_2^+$, it is appropriate to define $\Delta \varepsilon_\tau$ as
\begin{eqnarray}
\Delta \varepsilon_\tau = \varepsilon_\tau({\rm H}_2^+) - \frac{\varepsilon_\tau({\rm H_A} ) +\varepsilon_\tau({\rm H_B})}{2},
\end{eqnarray}
where ${\rm H_A}$ denotes the H atom at $z= -1.0$ and ${\rm H_B}$ at $z= 1.0$ (as noted above, more precise $R_e$ is 0.3\% shorter and the virial theorem holds only at the equilibrium position but we can practically consider $R_e$ to be 2.0\,bohrs). We plot this in the left panel of Fig.~\ref{fig:int}. From the definition, negative $\Delta \varepsilon_\tau$ (shown in blue) indicates stabilized regions and positive  $\Delta \varepsilon_\tau$ (shown in red) indicates destabilized regions. We see that stabilization mostly takes place between the H nuclei. The stabilized region slightly expands outside of the internuclear region and the zero surface of $\Delta \varepsilon_\tau$ is found to include the two nuclei. 

It is instructive to compare $\Delta \varepsilon_\tau$ with the differential electron density $\Delta n$, 
\begin{eqnarray}
\Delta n = n({\rm H}_2^+) - \frac{n({\rm H_A} ) +n({\rm H_B})}{2},
\end{eqnarray}
where $n = |\psi|^2$ is the ordinary electron density. This is plotted in the right panel of Fig.~\ref{fig:int}. We see that there is a region with increased electron density (shown in red) between the H nuclei and one with decreased electron density (shown in blue) outside the internuclear region. Compared with the map of $\Delta \varepsilon_\tau$, the region with increased electron density corresponds nicely to the stabilized region, which is consistent with a conventional picture of chemical bond formation. However, it should also be noticed that $\Delta \varepsilon_\tau$ and $\Delta n$ do not have one-to-one correspondence. For example, we see in Fig.~\ref{fig:int} that $\Delta \varepsilon_\tau$ has slightly larger zero surface that that of $\Delta n$. Moreover, integrating $\Delta n$ over the whole space gives zero whereas $\Delta \varepsilon_\tau$ gives the difference of binding energy between H$_2^+$ and H atom. 

In conclusion, we have calculated the stress tensor of the simplest molecule, H$_2^+$, and demonstrated its relevance to chemical bonding. It is not only the simplest but is also most accurate due to the existence of the exact wave functions. By using them, we are able to investigate the eigenvalue and eigenvector of the stress tensor and the tension vector field of both the ground state and the first excited state in a very accurate manner. It also makes possible the precise comparison between the interaction energy density, which tells the stabilized/destabilized regions, and conventional differential electron density with respect to the ground state. We believe that this calculation can be very useful as the most basic example which shows an important role of the electronic stress tensor for illustrating a chemical bond. Moreover, its exactness would serve as a benchmark to test validity of the calculation of the stress tensor which uses approximate wave functions. This is also very important since we can only obtain approximate wave functions for almost every molecular system. Therefore, future investigation should include how the stress tensor of H$_2^+$ deviates from the exact form as we change the level of approximation of the quantum chemistry calculation to derive H$_2^+$ wave functions. Such a study will solidify the basis of the stress tensor analysis of molecular systems and open up areas of further application.


\end{document}